\documentclass[preprint,tightenlines,eqsecnum,floats,aps,amsmath,amssymb,nofootinbib,prd,showpacs,superscriptaddress]{revtex4}  %one column
\usepackage{amsmath,amssymb,amsfonts}
\usepackage{graphicx}
\usepackage{enumerate}
\usepackage[colorlinks=true,linkcolor=blue,anchorcolor=blue,citecolor=blue,urlcolor=blue]{hyperref}
\usepackage[sort&compress]{natbib}
\usepackage{url}
\usepackage[T1]{fontenc}
\usepackage{mathptmx}
\usepackage{float}
\usepackage{booktabs}
\usepackage{CJK}
\usepackage{epstopdf}
\usepackage{tikz}
\usetikzlibrary{positioning, shapes.geometric, arrows}

\begin{document}

			\title{A new criterion for the existence of dark matter in neutron stars}
		%The Tidal Deformability and Moment of Inertia
		\author{Hongyi Sun}
		\affiliation{School of Physics and Optoelectronics, South China University of Technology, Guangzhou 510641, P.R. China}
		\author{ Dehua Wen\footnote{Corresponding author. wendehua@scut.edu.cn}}
		\affiliation{School of Physics and Optoelectronics, South China University of Technology, Guangzhou 510641, P.R. China}
		\date{\today}
		
		\begin{abstract}
			
			The tidal deformability and the radius of neutron stars are observables, which have been used to constrain the neutron star equation of state and explore the  composition in neutron stars. We investigated the radius and tidal deformability of dark matter admixed neutron stars (DANSs) by utilizing the two-fluid TOV equations.
			Assuming that the dark matter modeled as ideal fermi gas or self-interacting bosons, for a series of DANSs at a fixed mass, it is shown that there exists the DANSs with smaller normal matter radii but larger tidal deformabilities. This negative correlation does not exist in the normal neutron stars.
			In other words, if the observation finds that the neutron stars with a fixed mass exists such a situation, that is, having a smaller observed radius but a larger tidal deformability, it will indicate the existence of dark matter in neutron stars.
			In addition, the relevant neutron star observations can also be used to constrain the dark matter parameters.

			%	\begin{flushleft}
				%		%{Keywords: Dark matter admixed neutron stars; Equation of state; Radius; Tidal deformability}
				%	\end{flushleft}
		\end{abstract}

		\maketitle
		
		\section{Introduction}
		
		Despite years of researches and efforts, the essence of dark matter remains a mystery. There are several prospective particle candidates for the dark matter (such as Weakly Interacting Massive Particles (WIMPs), light bosons and sterile neutrinos\ \cite{Bertone2018A,Bertone2018B}).
		The interaction between the dark matter and the normal matter is mainly gravity, so a neutron star (one of the most compact objects in the universe) can capture a sizable amount of dark matter through its strong gravitational field.
		The accretion of dark matter into a neutron star can affect its macroscopic properties\ \cite{Ellis2018A,Sandin2009}, which can be used to extract information about the dark matter.
		In all, neutron stars provide a new way to investigate the nature of dark matter.

		Recently, the mass-radius measurements from the Neutron Star Interior Composition Explorer
		(NICER)  \cite{Riley2021,Miller2021,Riley2019, Miller2019} have yielded important constraints about the equations of state (EOSs) of dense nuclear matter\ \cite{Miao2022A,Li2021}.
		Besides, the observation of the binary neutron star merger, the GW170817 event\ \cite{Abbott2017}, has constrained the tidal deformability of a $M=1.4 ~M_{\odot}$ neutron star to a relatively small value ($\Lambda_{1.4}= 190^{+390}_{-120}$\ \cite{Abbott2018}), which favors a soft EOS.
		Numerous of investigations have used these observations to obtain  constraints on the properties of neutron stars and the parameters of EOS, such as the moment of inertia of neutron star \cite{I-Love-C,Jiang2020APJ,syang2022PRD,Silva2021PRL}, the sound speed \cite{Zhang2019,Sun2023,Kanakis-Pegios2021} and the symmetry energy \cite{LiYX2023,Zhang2023,Patra2023} of nuclear matter.
		
		Theoretically, the capture of dark matter will further influence the mass, radius and tidal deformability of neutron stars\ \cite{Deliyergiyev2019,Nelson2019,Rezaei2023,Liu2023}, and even lead to the appearance of supermassive neutron stars\ \cite{Lee2021}.
		The gravitational wave signal and the dynamics of the binary neutron star merger are also affected by the existence of dark matter\ \cite{Ellis2018,Ruter2018,Bezares2019,Horowitz2019}.
		The constraints on dark matter parameters obtained from the existing and future neutron star observations have also received extensive attentions\ \cite{Karkevandi2022,Bhattacharya2023,Banerjee2023}.
		In addition, the understanding of the interaction between the dark matter and the normal matter is still ambiguous.
		Except for the gravitational interaction, other interactions between the dark matter and the normal matter will soften the EOS, reducing the maximum mass of neutron stars\ \cite{Goldman2013} and the tidal deformability\ \cite{Lourenco2022,Das2020}.

		Furthermore,  the accumulation of dark matter in a neutron star will lead to the emergence of a new type of compact star, i.e., dark matter admixed neutron stars (DANSs)\ \cite{Thakur2023,Miao2022,Ivanytskyi2020,Shakeri2022,Giangrandi2022}.
		Similar to the normal neutron stars, if the gravitational interaction between the dark matter and the normal matter is only considered, it is found that the maximum mass of DANS corresponds to the beginning of the unstable sequence\ \cite{Leung2012}.
		For different types of dark matter candidates, such as fermionic dark matter\ \cite{Miao2022,Ivanytskyi2020}, self interacting bosonic dark matter\ \cite{Shakeri2022,Giangrandi2022}, and mirror dark matter\ \cite{Ciarcelluti2011,Sandin2009}, by adjusting the particle mass of dark matter and the interaction between the dark matter particles, DANSs with mass similar to the normal neutron stars can be obtained\ \cite{Leung2022,Rutherford2023}.
		In addition, if the dark matter mass fraction (i.e. the mass proportion of dark matter in DANSs) is large or the particle mass of dark matter is small, the dark matter radius tends to be greater than that of normal matter, forming a dark matter halo around the DANS\ \cite{Karkevandi2022}. Conversely, a small dark matter mass fraction and a large particle mass will lead to the occurance of a  dark matter core\ \cite{Miao2022}.
		Based on different processes of the DANS formation\ \cite{Ivanytskyi2020}, (for example, the remnant of the merger of a pure dark matter star\ \cite{Maselli2017,Gresham2019} and a neutron star may be a DANS with a large dark matter mass fraction\ \cite{Sandin2009}), DANSs with different dark matter mass fractions can be formed.

		In this work, the dark matter is modeled by self-interacting bosons\ \cite{Colpi1986} or ideal Fermi gas\ \cite{Oppenheimer1939,Leung2022}.
		Although other dark matter candidates, such as axions\ \cite{Lee2021,Desjacques2018}, self-interacting fermions\ \cite{Deliyergiyev2019,Miao2022}, and the mirror twin Higgs model\ \cite{Hippert2023}, may be more realistic, the two selected dark matter models are sufficient for a quantitative investigation of the  concerned properties of DANSs.
		Moreover, we only consider the gravitational interaction between the dark matter and the normal matter, so the two-fluid TOV equations\ \cite{Ciarcelluti2011} are  utilized to describe the DANSs.
		The influence of the dark matter on the structure and the properties of DANSs is
		investigated under different dark matter parameters.
		In addition, we focus on the change of the tidal deformability and the normal matter radius of DANSs, aiming to identify a way to prove the existence of dark matter in neutron stars by utilizing the uniqueness of DANS observables.
		
		This paper is organized as follows.
		In Sec. II, the basic formulas for the macroscopic properties of DANSs are briefly introduced. 	In Sec. III, the normal matter radius and the tidal deformability of DANSs with different dark matter parameters and different normal matter EOSs are presented. 	Finally, a summary is given in Sec. IV.

		\section{Basic formulas for the macroscopic properties of DANS}
		
			\subsection{Two-Fluid TOV Equations}
		
		In contrast to a nonrotating neutron stars with only one component, the individual components of a static neutron star with two fluids that only interact through gravity need to be calculated independently, i.e., using the two-fluid TOV equations\ \cite{Ciarcelluti2011,Giangrandi2022}
		\begin{align}
			\label{eq1}
			\frac{dp_{i}}{dr} =  -\frac{G\varepsilon(r)M(r)}{c^{2}r^{2}}
			\left(1+\frac{p_{i}(r)}{\varepsilon_{i}(r)}\right)\left(1+\frac{4\pi r^{3}p(r)}{M(r)c^{2}}\right)
			\left(1-\frac{2GM(r)}{c^{2}r}\right)^{-1},
		\end{align}
		\begin{equation}
			\label{eq2}
			\frac{dM_{i}}{dr} = \frac{4\pi r^{2}\varepsilon_{i}(r)}{c^{2}},
		\end{equation}
		where $i$ represents the two different components ($i = N$ or $D$ denotes the normal matter or the dark matter). $p_{i}(r)$, $\varepsilon_{i}(r)$, and $M_{i}(r)$ are the pressure, energy density, and mass of the two components at radius $r$, respectively.
		Variables without a subscript denote the sums of the two components  (i.e., $M(r) = M_{N}(r) + M_{D}(r)$, $p(r) = p_{N}(r) + p_{D}(r)$).
		The initial condition at the center of a neutron star are $M_{i}(0) = 0$ and  $p_{i}(0) = p_{c, i}$.
		The pressures of the two components will drop to 0 at different radii, then the mass and the radius of each component can be obtained.
		Due to the fact that the dark matter is electromagnetically dark, we focus on investigating the normal matter radius ($R_N$).

		\subsection{ Tidal Deformability}
		
		The tidal deformability of a normal neutron star characterizes the deformation of the neutron star due to the tidal effect created by a companion star.
		The details of the computation of the tidal deformability for nonrotating neutron stars can be found in Refs.\ \cite{Hinderer2010}. Here we only introduce the main equations. The tidal deformability is defined as
		\begin{equation}
			\lambda=\frac{2}{3}R^{5}k_2.
		\end{equation}
		The second (quadrupole) tidal Love number $k_2$ can be obtained from  \cite{Hinderer2010,Kanakis-Pegios2020}
		\begin{equation}
			\label{eq3}
			\begin{aligned}
				k_{2}=& \frac{8}{5}x^{5}(1-2x)^{2}(2-y_{R}+2x(y_{R}-1))
				(2x(6-3y_{R}+3x(5y_{R}-8))\\
				& +4x^{3}(13-11y_{R}+x(3y_{R}-2)+2x^{2}(1+y_{R}))\\
				& +3(1-2x)^{2}(2-y_{R}+2x(y_{R}-1))ln(1-2x))^{-1},
			\end{aligned}
		\end{equation}
		where $x=GM/Rc^{2}$ is the compactness of neutron stars, and  $y_{R}$ is determined by
		
		\begin{equation}
			r\frac{dy(r)}{dr}+y(r)^2+y(r)F(r)+r^2Q(r)=0,
			\label{eq4}
		\end{equation}
		with $F(r)$ and $Q(r)$ are functions of $M(r)$, $p(r)$ and $\varepsilon(r)$\ \cite{Kanakis-Pegios2020}
		\begin{equation}
			\label{eq5}
			F(r)=\left[1-\frac{4\pi r^{2}G}{c^4}(\varepsilon(r)-p(r))\right]\left(1-\frac{2M(r)G}{rc^2}\right)^{-1},
		\end{equation}
		\begin{equation}
			\label{eq6}
			\begin{aligned}
				r^2Q(r)=& \frac{4\pi r^{2}G}{c^4}\left[5\varepsilon(r)+9p(r)+\frac{\varepsilon(r)+p(r)}{\partial p(r)/\partial\varepsilon(r)}\right]\\
				&\times\left(1-\frac{2M(r)G}{rc^2}\right)^{-1}-6\left(1-\frac{2M(r)G}{rc^2}\right)^{-1}\\
				&-\frac{4M^{2}(r)G^{2}}{r^2c^4}\left(1+\frac{4\pi r^{3}p(r)}{M(r)c^2}\right)^{2}\left(1-\frac{2M(r) G}{rc^2}\right)^{-2},
			\end{aligned}
		\end{equation}
		then $y(r)=y_R$ at the radius $R$ of the neutron star is provided, and the dimensionless tidal deformability is given by
		\begin{equation}
			\label{eq7}
			\Lambda=\lambda\left(\frac{GM}{c^2}\right)^{-5}=\frac{2}{3}k_2\left(\frac{Rc^{2}}{GM}\right)^{5}.
		\end{equation}
		
		For a DANS with two fluids, Eq.(\ref{eq6}) should be modified. Specifically, the term with $\partial p(r)/\partial\varepsilon(r)$ should be changed to
		\begin{equation}
			\label{eq8}
			\frac{\varepsilon+p}{\partial p/\partial\varepsilon}\rightarrow\sum_{i}\frac{\varepsilon_{i}+p_{i}}{\partial p_{i}/\partial\varepsilon_{i}},
		\end{equation}
		and the energy density, pressure, and mass in Eqs. (\ref{eq5}) and (\ref{eq6}) should be replaced by the two components sums. Additionally, the total radius $R$ in Eq.\ (\ref{eq7}) is the larger one of the normal matter radius $R_N$ and the dark matter radius $R_D$.

		\subsection{Dark Matter Modeled as Self-interacting Bosons}

		Due to the absence of the degeneracy pressure in bosonic matter, it is necessary to assume the presence of self-interaction in bosonic dark matter to resist gravity.
		The EOS for the self-interacting bosonic dark matter\ \cite{Karkevandi2022,Colpi1986} can be written as
		\begin{equation}
			\label{eq14}
			p=\frac{4\rho_{0}c^2}{9}\left(\sqrt{1+\frac{3\rho}{4\rho_{0}}}-1\right)^{2}
		\end{equation}
		in the strong-coupling limit\ \cite{Colpi1986}, where $\rho_{0}=m_{\chi, b}^{4} / 4\lambda\hbar^{3}c^{5}$, and $m_{\chi, b}$, $\rho$, and $\lambda$ are the particle mass, density, and coupling constant of the self-interacting bosonic dark matter.
		From Eq. (\ref{eq14}), it is evident that increasing $m_{\chi, b}$ and decreasing $\lambda$ have similar effects on the EOS of bosonic dark matter, as the effective parameter is ${m_{\chi, b}^{4}}/{\lambda}$.
		In the following, the coupling constant is fixed to $\lambda = \pi$ (which satisfies the strong-coupling limit) and we only consider the impact of $m_{\chi, b}$ on DANSs.

		\subsection{Dark Matter Modeled as Ideal Fermi Gas} Moreover, we assume the dark matter to be made of ideal Fermi gas and neglect finite temperature effects.
		Although this model is considered relatively simplistic, it still provides us with quantitative insights into the properties of DANS\ \cite{Leung2022}.
		Assuming there is only one type of dark matter particle with spin $\frac{1}{2}$, the ideal Fermi gas EOS\ \cite{Oppenheimer1939,Leung2022} at zero temperature is
		\begin{equation}
			\label{eq9}
			\varepsilon=K(\textrm{sinh}t-t),
		\end{equation}
		\begin{equation}
			\label{eq10}
			p=\frac{K}{3}\left(\textrm{sinh}t-8\textrm{sinh}\frac{t}{2}+3t\right),
		\end{equation}
		with
		\begin{equation}
			\label{eq11}
			K=\frac{\pi m_{\chi, f}^{4}}{32\pi^{3}(\hbar c)^{3}},
		\end{equation}
		\begin{equation}
			\label{eq12}
			t=4\textrm{ln}(y+\sqrt{1+y^2}),
		\end{equation}
		where $y$ is a variable related to the number density $n$
		\begin{equation}
			\label{eq13}
			y=\left(\frac{3\pi^{2}(\hbar c)^{3}n}{m_{\chi, f}^{3}}\right)^{1/3},
		\end{equation}
		and $m_{\chi, f}$ is the particle mass, which is also the only adjustable parameter of the ideal Fermi gas EOS.

		\subsection{EOS of Normal Matter}

		In order to meet the constraints of the GW170817 event and the NICER observational data, we adopt the APR3 EOS (soft)\ \cite{APR3} to represent the normal matter EOS.
		For comparison, we also presented the results obtained using the DDME2 EOS (stiff)\ \cite{DDME2} as the normal matter EOS, and show the influences of different normal matter EOSs on the macroscopic properties of DANSs.
		In addition, if the dark matter EOSs is sufficiently stiff, the maximum mass of DANSs constructed with these normal matter EOSs may exceed that of normal neutron stars, and the mass-radius (M-R) relation of DANSs is no longer a curve but extends to a M-R plane\ \cite{Hippert2023,Li2012,Zollner2023}.

		\section{ Relations of Tidal Deformability and normal matter radius of DANS}	
		\subsection{Macroscopic Properties of DANSs}	
		
\begin{figure}[h]
			\centering
			\includegraphics[width=1\textwidth]{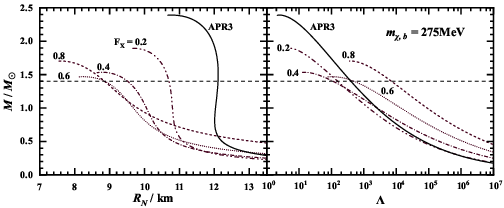}
			\caption{The influences of different dark matter mass fractions ($F_X=M_{D}/M$) on the $M-R_N$ (left) and $M-\Lambda$ relations (right) of DANSs (dashed lines), where the normal matter EOS is the APR3 EOS and the particle mass of self-interacting bosonic dark matter is $m_{\chi, b}=200$MeV. The $M-R_N$ and $M-\Lambda$ relations of normal neutron stars (solid lines) calculated by the APR3 EOS are also shown.}
			\label{fig:1}
		\end{figure}

		Based on the two fluid TOV equations, the relations of the mass and the normal matter radius ($M-R_N$) of DANSs, as well as the  mass-tidal deformability ($M-\Lambda$) relations, can be obtained by providing a normal matter EOS and a  dark matter EOS.
		In Fig.\ \ref{fig:1}, the influences of different dark matter mass fractions ($F_X=M_{D}/M$) on the $M-R_N$ (left panel) and $M-\Lambda$ (right panel) relations of DANSs are shown, where the normal matter EOS is the APR3 EOS and the particle mass of self-interacting bosonic dark matter is $m_{\chi, b}=275$MeV.
		The $M-R_N$ and $M-\Lambda$ relations of normal neutron stars (without the dark matter) are also displayed for comparison.
		Here we focus on the normal matter radius $R_N$.
		The dark matter radius $R_D$ can be hardly detected because the dark matter is electromagnetically dark, but the $R_D$ can have a significant impact on the tidal deformability of DANSs\ \cite{Nelson2019,Routaray2023}.
		
		From the left panel of Fig.\ \ref{fig:1}, it is shown that the $M-R_N$ relations of DANSs will extend to a series of curves (dashed lines) instead of one curve (solid line) with the change of dark matter mass fraction $F_X$.
		If the $F_X$ takes continuous values instead of discrete values, these $M-R_N$ curves will form a $M-R_N$ plane, which means there are multiple DANSs with the same mass but different normal matter radii.
		This feature of DANSs is similar to that of twin stars\ \cite{Blaschke2020}.
		However, in the $M-R_N$ relations containing twin stars, only two different radii are allowed for the normal neutron stars with the same mass\ \cite{Alford2016}, thus the twin stars can be distinguished from the DANSs through observations.
		Due to the wide range of the $R_N$ at a fixed stellar mass of DANSs, induced by the variation of the $F_X$, the $M-R_N$ relations basically satisfy the constraints of existing observational data, and it is difficult to prove the absence of dark matter in neutron stars only through the existing observations, but the amount of dark matter in the DANSs can be constrained through these observations\ \cite{Ivanytskyi2020}.
		Additionally, as the $F_X$ increases, the maximum mass of DANS will decreases firstly and then increases, and the $R_N$ will decreases\ \cite{Leung2022,Li2012} (but for large $F_X$, the $R_N$ of low mass DANSs may increase).
		
		Unlike the normal matter radius $R_N$, the tidal deformability $\Lambda$ of DANSs in the right panel of Fig.\ \ref{fig:1} decreases firstly and then increases with the increase of the dark matter mass fraction $F_X$.
		Moreover, when the $F_X$ is high ($\sim0.8$), the $\Lambda$ of DANSs will become very large\ \cite{Lee2021} (for example, the $\Lambda$ of DANSs with a mass of $1.4M_{\odot}$ can reach $\Lambda_{1.4}\sim8000$) compared to that of normal neutron stars (for the APR3 EOS, $\Lambda_{1.4}\sim400$).
		This large $\Lambda$ of DANSs is caused by the large dark matter radius $R_D$.
		For a higher $F_X$, $R_D > R_N$ (see the lower panel of Fig.\ \ref{fig:3}), so even if the $R_N$ is small (see the left panel of Fig.\ \ref{fig:1}), the total radius $R$ is determined by the larger $R_D$, and it will result in a large $\Lambda$ of a DANS (see Eq.(\ref{eq7})) greater than the $\Lambda$ of a normal neutron star.
		
\begin{figure}[h]
			\centering
			\includegraphics[width=1\textwidth]{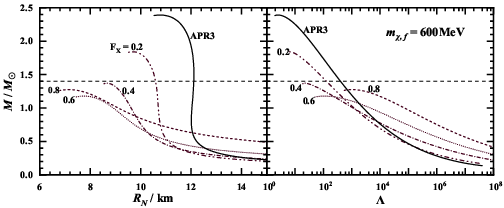}
			\caption{The same as Fig.\ \ref{fig:1}, but the dark matter is modeled by ideal Fermi gas and the particle mass is $m_{\chi, f}=600$MeV.}
			\label{fig:2}
		\end{figure}

		For the dark matter modeled as the ideal Fermi gas, the influence of the dark matter mass fraction $F_X$ on the normal matter radius $R_N$ and the tidal deformability $\Lambda$ of DANSs are shown in Fig.\ \ref{fig:2}.
		Similar to Fig.\ \ref{fig:1}, the $M-R_N$ and $M-\Lambda$ relations of DANSs have also extended to a series of curves.
		However, compared to the self-interacting bosonic dark matter, a larger particle mass is need in the ideal Fermi gas EOS to support DANSs with similar mass.
		
		\subsection{Negative correlation between tidal deformability and normal matter radius}
\begin{figure}[h]
			\centering
			\includegraphics[width=0.5\textwidth]{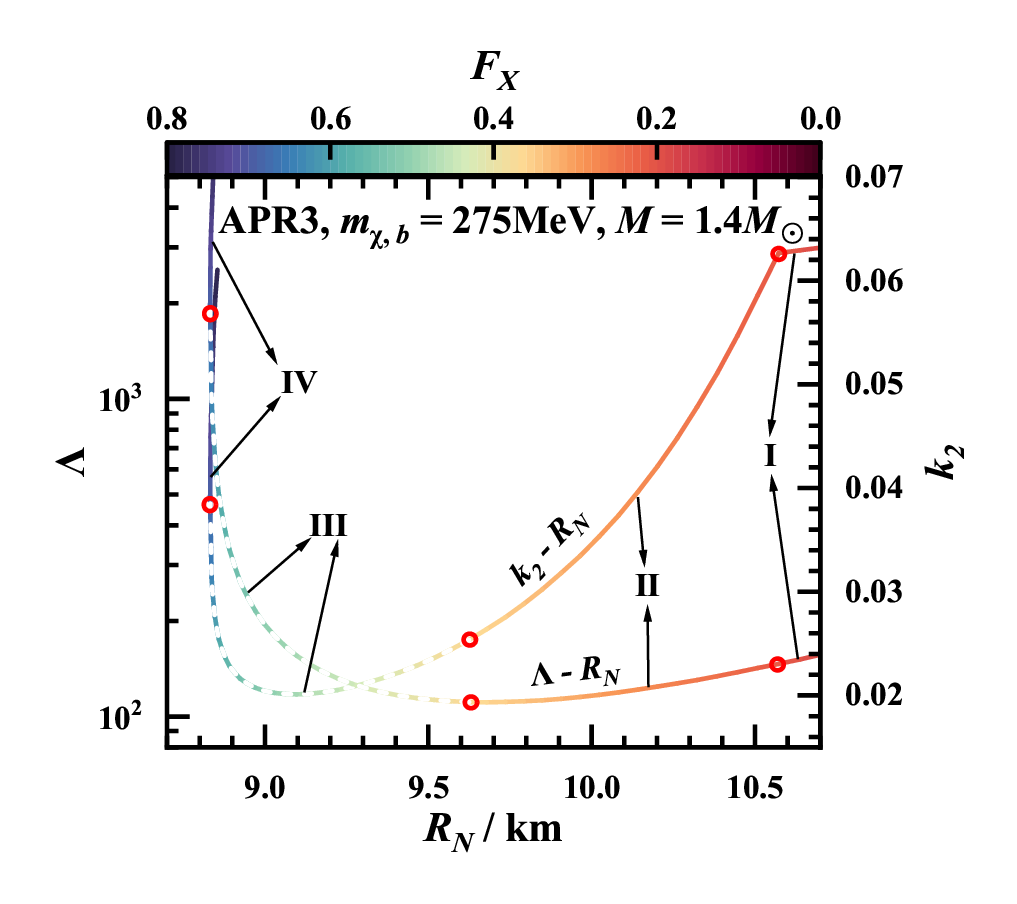}
			\includegraphics[width=0.5\textwidth]{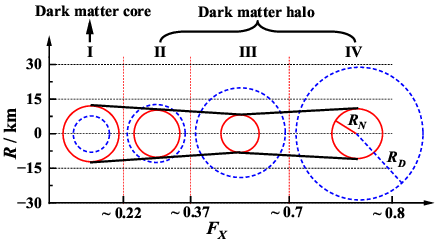}
			\caption{Upper panel: The $k_2-R_N$ and $\Lambda-R_N$ relations of DANSs with a fixed stellar mass ($1.4M_{\odot}$) at different dark matter mass fractions $F_X$.
				The dashed lines labeled with III represent the range of $F_X$ corresponding to the negative correlation between the tidal deformability $\Lambda$ and the normal matter radius $R_N$, while the solid lines (I, II, and IV) represent the positive correlation between the $\Lambda$ and the $R_N$.
				Lower panel: The relative size of the $R_N$ (red solid circle) and the $R_D$ (blue dashed circle) of DANSs ($1.4M_{\odot}$) at different $F_X$ values (I, II, III, and VI).
				The normal matter EOS and the dark matter parameters adopt the same as those used in Fig.\ \ref{fig:1}.
			}
			\label{fig:3}
		\end{figure}

		To illustrate the effect of dark matter on the normal matter radius $R_N$ and the (dimensionless) tidal deformability $\Lambda$ of DANSs intuitively, the upper panel of Fig.\ \ref{fig:3} shows the relation of $R_N$ and $\Lambda$ of  DANSs with a fixed stellar mass ($1.4M_{\odot}$) at different dark matter mass fractions $F_X$.
		Here we only show the result of self-interacting bosonic dark matter, and the result of ideal Fermi gas is similar to that.
		From the upper panel of Fig.\ \ref{fig:3}, it can be seen that with the increase of $F_X$, both the $R_N$ and the $\Lambda$ initially decrease and then increase. However, their decreases or increases are not simultaneous.
		That is to say, in a specific range of $F_X$ (indicated by the dashed line labeled III in the upper panel of Fig.\ \ref{fig:3}), the $\Lambda$ increases while the $R_N$ decreases.
		As a comparison, this phenomenon does not exist in twin stars.
		Although the mass of twin stars are the same, the one with a smaller radius must has a smaller tidal deformability\ \cite{Blaschke2020}.
		Therefore, the negative correlation between the tidal deformability $\Lambda$ and the normal matter radius $R_N$ of the DANSs at a fixed stellar mass  may served as a clue to prove the presence of dark matter in neutron stars.
		
		To further investigate the negative correlation between  the tidal deformability $\Lambda$ and the normal matter radius $R_N$ of DANSs, the $k_2-R_N$ relation of DANSs with a mass of $1.4M_{\odot}$ at different $F_X$ is also shown in the upper panel of Fig.\ \ref{fig:3} (where $k_2$ is the tidal Love number), and the relative size of the $R_N$ and the $R_D$ of those DANSs is presented in the lower panel of Fig.\ \ref{fig:3}.
		For DANSs with a relatively small $F_X$ (labeled with I and II in Fig.\ \ref{fig:3}), an increase of $F_X$ leads to a decrease of $R_N$, $k_2$ and $\Lambda$. Although in the stage II (beginning with the kink of the $k_2-R_N$ curve) the total radius $R = R_D>R_N$ (i.e., dark matter halo) increases with the increase of $F_X$, the decreasing $k_2$ dominates the decrease of tidal deformability $\Lambda$ (see Eq.\ (\ref{eq7})).
		Conversely, the increase of $R=R_D$ will dominate as the $F_X$ further increases (III in Fig.\ \ref{fig:3}), so the $\Lambda$ will increase while the $R_N$ remains decreasing.
		This range of $F_X$ corresponds to the negative correlation between the tidal deformability $\Lambda$ and the normal matter radius $R_N$.
		When the dark matter accounts for a significant portion in DANSs (VI in Fig.\ \ref{fig:3}), the $\Lambda$ and the $R_N$ show a positive correlation again.
		It should be noted that for the DANSs with different masses, the specific  range of $F_X$ corresponding to the negative correlation between the tidal deformability $\Lambda$ and the normal matter radius $R_N$ varies (see Fig.\ \ref{fig:4}).
		
		\subsection{Influences of dark matter parameters and normal matter EOSs}
		
\begin{figure}[ht]
			\centering
			\includegraphics[width=0.8\textwidth]{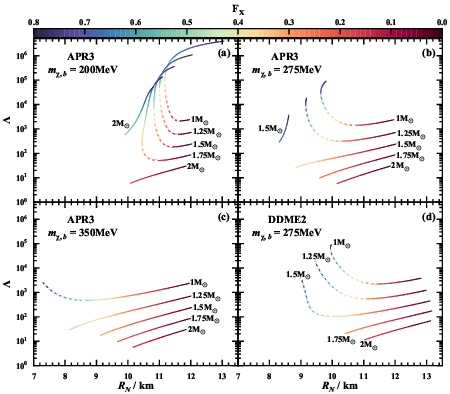}
			\caption{The effects of the particle mass $m_{\chi, b}$ of self-interacting dark matter and the normal matter EOSs on the relations of normal matter radius $R_N$ and tidal deformability $\Lambda$ of DANSs (with masses of 1, 1.25, 1.5, 1.75, and 2 $M_{\odot}$) at different dark matter mass fractions $F_X$.
				In panels (a), (b), and (c), the $m_{\chi, b}$ adopts 200, 275, and 350 MeV, and the normal matter is described by the APR3 EOS.
				In panel (d), the $m_{\chi, b}$ adopts 275 MeV, and the normal matter is described by the DDME2 EOS.
				The dashed and solid lines are similar to those in Fig.\ \ref{fig:3}.}
			\label{fig:4}
		\end{figure}

		%$\textit{Influences ~of ~dark~ matter~ parameters~ and~ normal~ matter~ EOSs}.$
		The normal matter radius $R_N$ and tidal deformablity $\Lambda$ of DANSs are also influenced by the particle mass of dark matter and the normal matter EOSs.
		In order to investigate the generality of the negative correlation between the $\Lambda$ and the $R_N$, in Fig.\ \ref*{fig:4}, the three panels, (a), (b), and (c), show the effect of the particle mass $m_{\chi, b}$ of self-interacting bosonic dark matter on the $\Lambda$ and $R_N$, and the panels (b) and (d) show the effect of the normal matter EOSs.
		The discontinuous solid lines (such as the $2M_{\odot}$ DANSs in the panel (a)) indicate that for some ranges of $F_X$, the given EOSs of normal matter and dark matter can not support the DANSs with a relatively high mass ($2M_{\odot}$).
		
		The three panels (a), (b), and (c) of Fig.\ \ref{fig:4} show that an increase of $m_{\chi, b}$ will induce a decrease of tidal deformability $\Lambda$ and normal matter radius $R_N$ of the DANSs with a large $F_X$.
		For a larger $m_{\chi, b}$, the self-interacting bosonic dark matter tends to accumulate within the DANSs to form a dark matter core rather than a dark matter halo, and the heavier core has a stronger gravitational effect on the normal matter, resulting in a decrease of $R_N$.
		Similarly, the larger $m_{\chi, b}$ also lead to a smaller total radius $R$ of DANSs, which induces a smaller $\Lambda$.
		Furthermore,  the increase of $m_{\chi, b}$ will induce the negative correlation between the tidal deformability $\Lambda$ and the normal matter radius $R_N$ (dashed lines) to occur at a lower DANS mass.
		For $m_{\chi, b} =$ 200 MeV, this negative correlation of DANSs with a mass of 1.75 $M_{\odot}$ is displayed in the panel (a) of Fig.\ \ref*{fig:4}, but for $m_{\chi, b} =$ 275 MeV, that of DANSs only occurs when the mass of DANSs is smaller than 1.5 $M_{\odot}$ (panel (b) of Fig.\ \ref*{fig:4}).
		If the $m_{\chi, b}$ further increases ($m_{\chi, b} >$ 350 MeV), then only the DANSs with a mass around 1 $M_{\odot}$ or less than that are likely to show the negative correlation between the tidal deformability $\Lambda$ and the normal matter radius $R_N$.
		Conversely, a smaller particle mass ($m_{\chi, b} <$ 200 MeV) will lead to a less obvious decrease in the $R_N$ of DANSs with this negative correlation during the increase of $\Lambda$, making it difficult to distinguish such DNASs through the normal matter radius $R_N$.
		Therefore, we only investigate the macroscopic properties of DANSs when the $m_{\chi, b}$ is at around 300 MeV.	
		Additionally, a stiffer normal matter EOS (panel (d) of Fig.\ \ref{fig:4}) can cause this negative correlation to occur when the DANSs have a larger mass (1.5 $M_{\odot}$), but the effect of normal matter EOSs is not as significant as the effect of varying $m_{\chi, b}$.
		Overall, for the self-interacting bosonic dark matter, the negative correlation between the tidal deformability  and the normal matter radius of DANSs is universal when changing the particle mass of dark matter and the normal matter EOS.
		Furthermore, it is important to emphasize that the effective parameter in Eq.(\ref{eq14}) is $m_{\chi, b}^4/\lambda$, so the effect of decreasing $\lambda$ is the same as the effect of increasing $m_{\chi, b}$.
		If the $\lambda$ is changed, the tidal deformability and the normal matter radius of DANSs at a fixed $m_{\chi, b}$ will also change.

		\begin{figure}[h]
			\centering
			\includegraphics[width=0.8\textwidth]{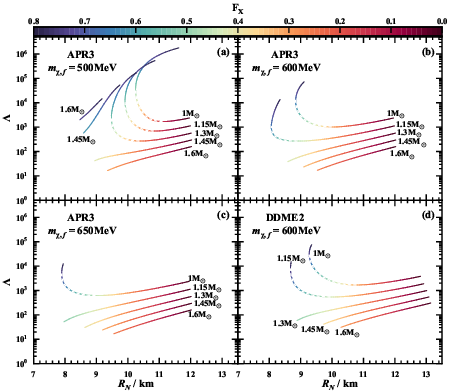}
			\caption{The same as Fig.\ \ref{fig:4}, but the dark matter is modeled by the ideal Fermi Gas, and the particle mass $m_{\chi, f}$ of fermionic dark matter adopts 500, 600, and 650 MeV. The masses of DANSs are 1, 1.15, 1.3, 1.45, and 1.6 $M_{\odot}$.}
			\label{fig:5}
		\end{figure}
		
		Similarly, figure\ \ref{fig:5} shows the effects of varying the particle mass $m_{\chi, f}$ of fermionic dark matter and the normal matter EOS on the tidal deformability $\Lambda$ and the normal matter radius $R_N$ of DANSs.
		For the dark matter modeled as ideal Fermi gas, a larger particle mass is needed to support DANSs with similar stellar mass compared with the situation in the case of self-interacting bosoninc dark matter.
		The negative correlation between the tidal deformability $\Lambda$ and the normal matter radius $R_N$ of DANSs is also presented for the fermionic dark matter, as shown in Fig.\ \ref{fig:5}.
		The increase of $m_{\chi, f}$ also causes this negative correlation to occur at a lower DANS mass, whereas the effect of modifying the normal matter EOS on this negative correlation is not obvious.
		As a result, regardless of the types of dark matter (self-interacting bosons or ideal Fermi gas), if there are a series of neutron stars with the same or extremely similar mass in future observation, and the ones with smaller radii have larger tidal deformabilities, it is very likely to imply the existence of dark matter in the neutron stars.
		Moreover, if the stellar mass of this series of neutron stars is greater than a certain value (for example, $M > 1.25M_{\odot}$), then it is also possible to constrain the parameters of dark matter (for the self-interacting bosonic dark matter, $m_{\chi, b} <$ 350 MeV when $\lambda = \pi$, and for the dark matter modeled as ideal Fermi gas, the $m_{\chi, f}$ is constrained to $m_{\chi, f} <$ 600 MeV).

		\section{summary}
		%$\textit{Summary}.$
		It is shown that there is a series of DANSs with the same mass but different dark matter mass fractions, and the ones with smaller normal matter radii have larger tidal deformabilities, i.e., the negative correlation between the tidal deformability and the normal matter radius.
		This negative correlation does not exist in other types of neutron stars (such as twin stars\ \cite{Blaschke2020}, quark stars\ \cite{Albino2021}).
		If observations confirm that there is a series of neutron stars with the same or very similar mass, and the ones with smaller radii have greater tidal deformabilities, it may indicate the existence of dark matter in the neutron stars.
		The interaction among the dark matter particles may be more complex in reality\ \cite{Deliyergiyev2019,Lee2021,Routaray2023}, the negative correlation between the tidal deformability and the normal matter radius of DANSs is believed to exist even when the dark matter model is changed.
		In addition, the negative correlation between the tidal deformability and the normal matter radius of DANSs also provide a way to constrain the dark matter parameters.

		\section{acknowledgement}
		This work is supported by NSFC (Grant No. 12375144, 11975101) and Guangdong Natural Science Foundation (Grants No. 2022A1515011552 and No. 2020A151501820).

	\end{document}